\documentstyle[12pt,twoside,fleqn,epsfig,espcrc1]{article} 
\begin{document} 

\title{$\Lambda\Lambda$ hypernuclei and stranger systems\thanks
{Supported in part by the Israel Science Foundation, Jerusalem, 
grant 01/131.}} 

\author{Avraham Gal\address{Racah Institute of Physics, 
The Hebrew University, Jerusalem 91904, Israel}} 

\maketitle

\begin{abstract}

Recent experiments on production of $\Lambda\Lambda$ Hypernuclei 
have stimulated renewed interest in extracting the $\Lambda\Lambda$ 
interaction from the few events identified since the inception of 
this field forty years ago. Few-body calculations relating to this 
issue are reviewed, particularly with respect to the possibility that 
$A=4$ marks the onset of $\Lambda\Lambda$ binding to nuclei. 
The Nijmegen soft-core model potentials NSC97 qualitatively 
agree with the strength of the $\Lambda\Lambda$ interaction deduced 
from the newly determined binding energy of $_{\Lambda\Lambda}^{~~6}$He. 
Applying the extended NSC97 model to stranger nuclear systems suggests 
that $A=6$ marks the onset of $\Xi$ binding, with a particle stable 
$_{\Lambda\Xi}^{~~6}$He, and that strange hadronic matter is robustly bound. 
\end{abstract} 

%\PACS{21.80.+a, 11.80.Jy, 21.10.Dr, 21.45.+v} 

\section{INTRODUCTION} 

Until 2001 only three candidates existed for $\Lambda\Lambda$ 
hypernuclei observed in emulsion experiments \cite{Dan63,Pro66,Aok91}. 
The $\Lambda\Lambda$ binding energies 
deduced from these emulsion events indicated that the $\Lambda\Lambda$ 
interaction is strongly attractive in the $^{1}S_0$ channel 
\cite{DDF89,DMGD91,YTI91}, with a $\Lambda\Lambda$ pairing energy 
$\Delta B_{\Lambda\Lambda} \sim 4.5$ MeV, although it had been 
realized \cite{BUC84,WTB86} that the binding energies of 
$_{\Lambda\Lambda}^{~10}$Be \cite{Dan63} and $_{\Lambda\Lambda}^{~~6}$He 
\cite{Pro66} are inconsistent with each other. This outlook has undergone 
an important change following the very recent report by the KEK 
hybrid-emulsion experiment E373 of a well-established new candidate 
\cite{Tak01} for $_{\Lambda\Lambda}^{~~6}$He, with binding energy 
($\Delta B_{\Lambda\Lambda} \sim 1$ MeV) substantially lower than 
that deduced from the older, dubious event \cite{Pro66}. 
Furthermore, there are also indications from the 
AGS experiment E906 for the production of light $\Lambda\Lambda$ 
hypernuclei \cite{Ahn01}, perhaps as light even as 
$_{\Lambda\Lambda}^{~~4}$H, in the $(K^-,K^+)$ reaction on $^9$Be. 

Since data on hyperon-nucleon ($YN$) and hyperon-hyperon ($YY$) 
interactions are scarce or even not readily available from laboratory 
experiments, the study of multistrange systems provides a fairly 
exclusive test of microscopic models for the baryon-baryon ($BB$) 
interaction. The Nijmegen group has constructed over the years 
a number of one-boson-exchange (OBE) models (reviewed by Rijken in 
Ref.\cite{Rij01} and in these proceedings) 
for the $BB$ interaction using SU(3)-flavor symmetry to relate coupling 
constants and phenomenological short-distance hard or soft cores. 
In all of these rather different $BB$ interaction models only 
35 $YN$ low-energy, generally imprecise data points serve the purpose 
of steering phenomenologically the extrapolation from the $NN$ sector, 
which relies on thousands of data points, into the strange 
$YN$ and $YY$ sectors. It is therefore of utmost importance to confront 
these models with the new $\Lambda\Lambda$ hypernuclear data in order 
to provide meaningful constraints on the extrapolation to strangeness 
$S = -2$ and beyond. 

\begin{figure} 
\begin{minipage}[t]{75mm} 
\epsfig{file=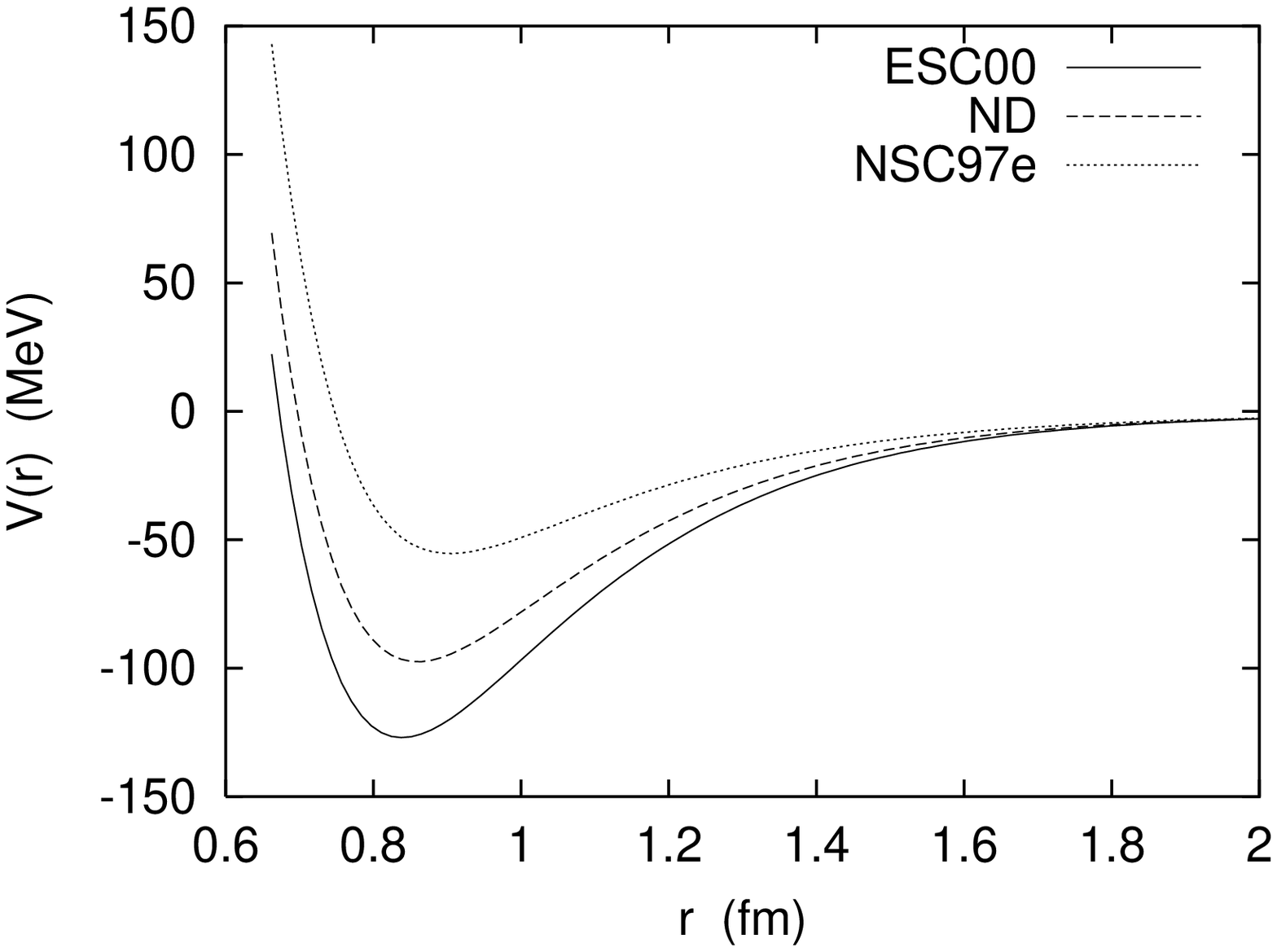,height=80mm,width=75mm} 
\caption{Nijmegen OBE phase-equivalent soft-core $\Lambda\Lambda$ 
potentials \cite{FGa02b}.} 
\label{fig:llpot} 
\end{minipage} 
\hspace{\fill} 
\begin{minipage}[t]{75mm} 
\epsfig{file=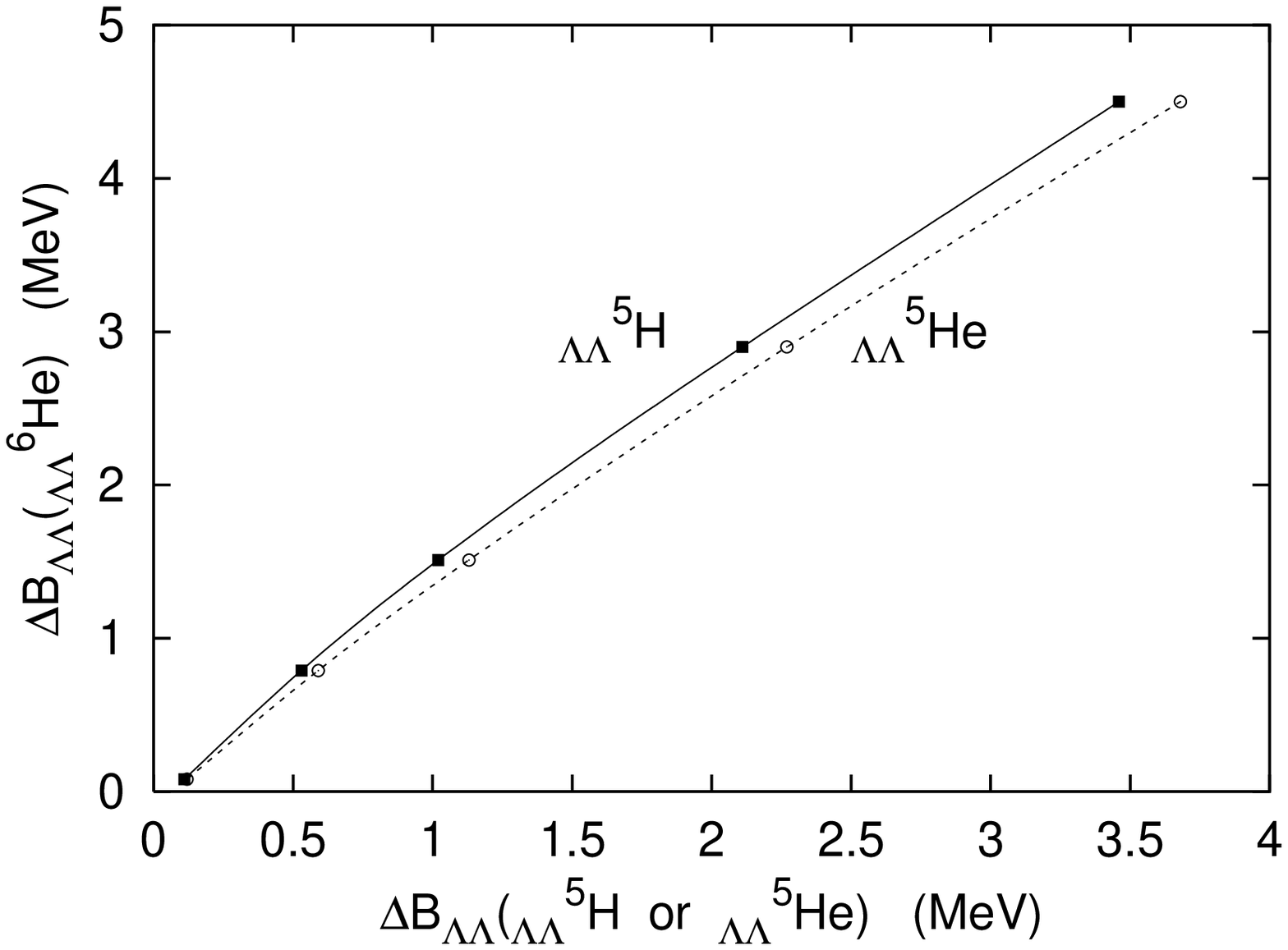,height=80mm,width=75mm} 
\caption{$s$-wave Faddeev calculations \cite{FGa02b} of 
$\Delta B_{\Lambda\Lambda}(_{\Lambda\Lambda}^{~~6}\rm He)$ 
vs. $\Delta B_{\Lambda\Lambda}(_{\Lambda\Lambda}^{~~5}\rm H,~
_{\Lambda\Lambda}^{~~5}\rm He)$.} 
\label{fig:e6e5} 
\end{minipage} 
\end{figure} 

The $YN$ and $YY$ $s$-wave interaction input potentials to the structure 
calculations here reviewed often consist of combinations of Gaussians 
with different ranges, such as to make these single-channel potentials 
phase equivalent to the Nijmegen OBE-model coupled-channel potentials. 
Of the several $\Lambda\Lambda$ potentials due to Nijmegen models which 
are shown in Fig.\ref{fig:llpot}, NSC97e is the weakest one, of the order 
of magnitude required to reproduce 
$\Delta B_{\Lambda\Lambda}(_{\Lambda\Lambda}^{~~6}$He). 
The $\Lambda\Lambda$ interaction is fairly weak for all six versions 
($a$)-($f$) of the Nijmegen soft-core model NSC97 \cite{SRi99}, 
and versions $e$ and $f$ provide a reasonable description of 
single-$\Lambda$ hypernuclei \cite{RSY99}. 

\section{$\Lambda\Lambda$ HYPERNUCLEI} 

In this section I will review topical theoretical work on some of the 
light $\Lambda\Lambda$ hypernuclear species connected to old and to 
new experiments. The anticipated existence of $_{\Lambda\Lambda}^{~~6}$He, 
now solidly established also experimentally \cite{Tak01}, leads one to 
enquire where the onset of $\Lambda\Lambda$ binding occurs. It was argued 
long ago that the three-body $\Lambda\Lambda N$ system is unbound 
\cite{THe65}, and hence I will concentrate on the $A=4,5$ $\Lambda\Lambda$ 
hypernuclear systems. Among the few heavier species reported todate, 
$_{\Lambda\Lambda}^{~10}$Be will be discussed briefly. 

\subsection
{$_{\Lambda\Lambda}^{~~5}${\rm H} - $_{\Lambda\Lambda}^{~~5}${\rm He}} 

Figure \ref{fig:e6e5} demonstrates a nearly linear correlation between 
Faddeev-calculated values of 
$\Delta B_{\Lambda\Lambda}(_{\Lambda\Lambda}^{~~6}$He) and 
$\Delta B_{\Lambda\Lambda}(_{\Lambda\Lambda}^{~~5}$H, 
$_{\Lambda\Lambda}^{~~5}$He), using several $\Lambda\Lambda$ interactions 
which include (the lowest-left point) $V_{\Lambda\Lambda} = 0$ \cite{FGa02b}. 
Here 
\begin{equation} 
\label{eq:delB} 
\Delta B_{\Lambda\Lambda} (^{~A}_{\Lambda \Lambda}Z) 
= B_{\Lambda\Lambda} (^{~A}_{\Lambda \Lambda}Z) 
- 2{\bar B}_{\Lambda} (^{(A-1)}_{~~\Lambda}Z)\;, 
\end{equation} 
where $B_{\Lambda\Lambda} (^{~A}_{\Lambda \Lambda}Z)$ is the $\Lambda\Lambda$ 
binding energy of the hypernucleus $^{~A}_{\Lambda \Lambda}Z$ and 
${\bar B}_{\Lambda} (^{(A-1)}_{~~\Lambda}Z)$ is the (2$J$+1)-average of 
$B_{\Lambda}$ values for the $^{(A-1)}_{~~\Lambda}Z$ hypernuclear core levels. 
$\Delta B_{\Lambda\Lambda}$ increases monotonically with the strength of 
$V_{\Lambda\Lambda}$, starting in approximately zero as 
$V_{\Lambda\Lambda} \rightarrow 0$, which is a general 
feature of three-body models such as the $\alpha\Lambda\Lambda$, 
$^3$H$\Lambda\Lambda$ and $^3$He$\Lambda\Lambda$ models used in these 
$s$-wave Faddeev calculations \cite{FGa02b}, and also as shown below for 
$d\Lambda\Lambda$ $s$-wave Faddeev calculations \cite{FGa02c}. The $I = 1/2$ 
$^{~~5}_{\Lambda\Lambda}$H - $^{~~5}_{\Lambda\Lambda}$He hypernuclei are 
then found to be particle stable for {\it all} the $\Lambda\Lambda$ attractive 
potentials here used. This conclusion holds also when the $s$-wave 
approximation is relaxed \cite{FGS03}.

\begin{figure} 
\begin{minipage}[t]{75mm} 
\epsfig{file=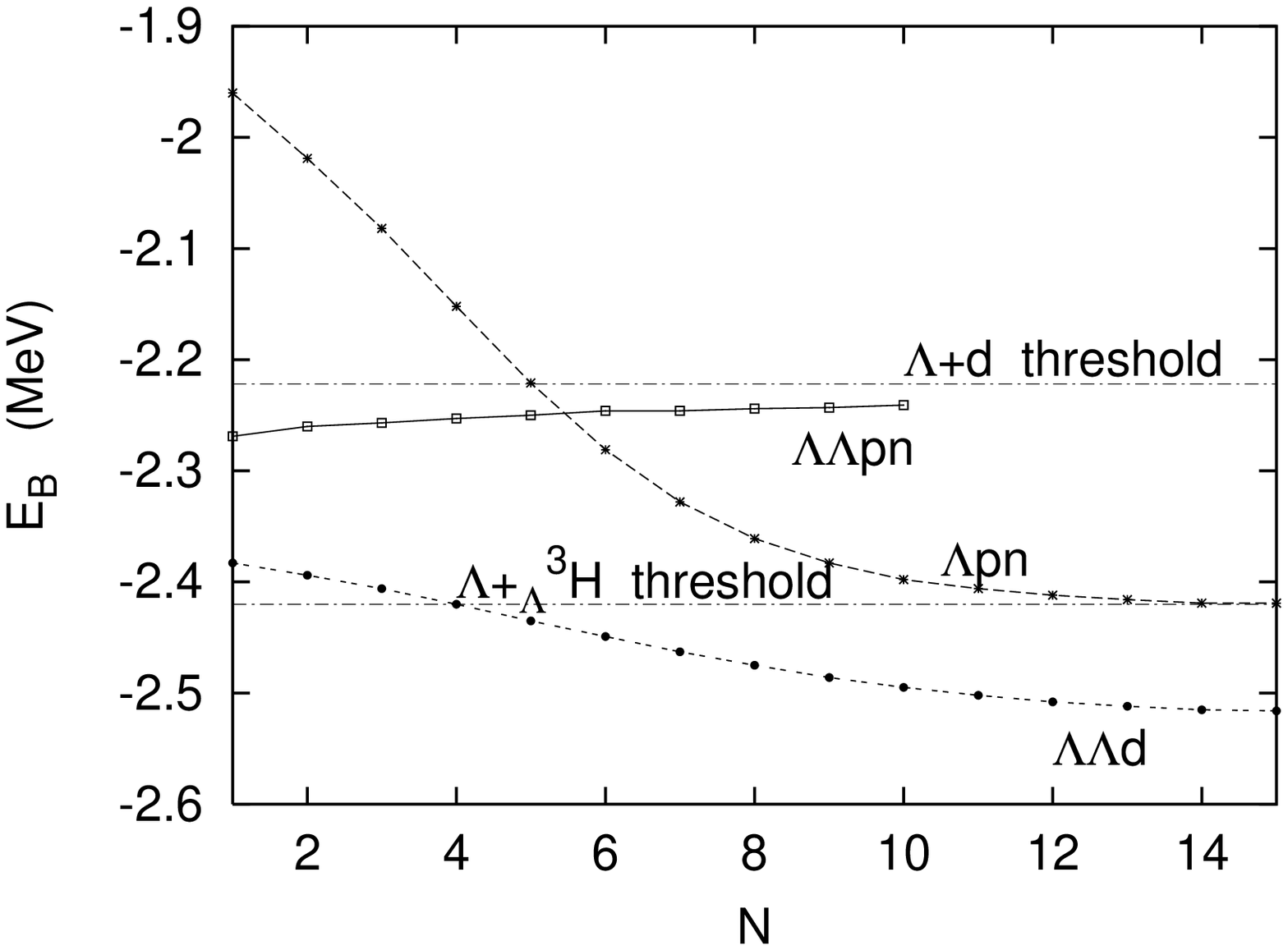,height=80mm,width=75mm} 
\caption{$s$-wave FY calculations \cite{FGa02c} for 
$\Lambda pn$, $\Lambda\Lambda d$ and $\Lambda\Lambda pn$.} 
\label{fig:BE} 
\end{minipage} 
\hspace{\fill} 
\begin{minipage}[t]{75mm} 
\epsfig{file=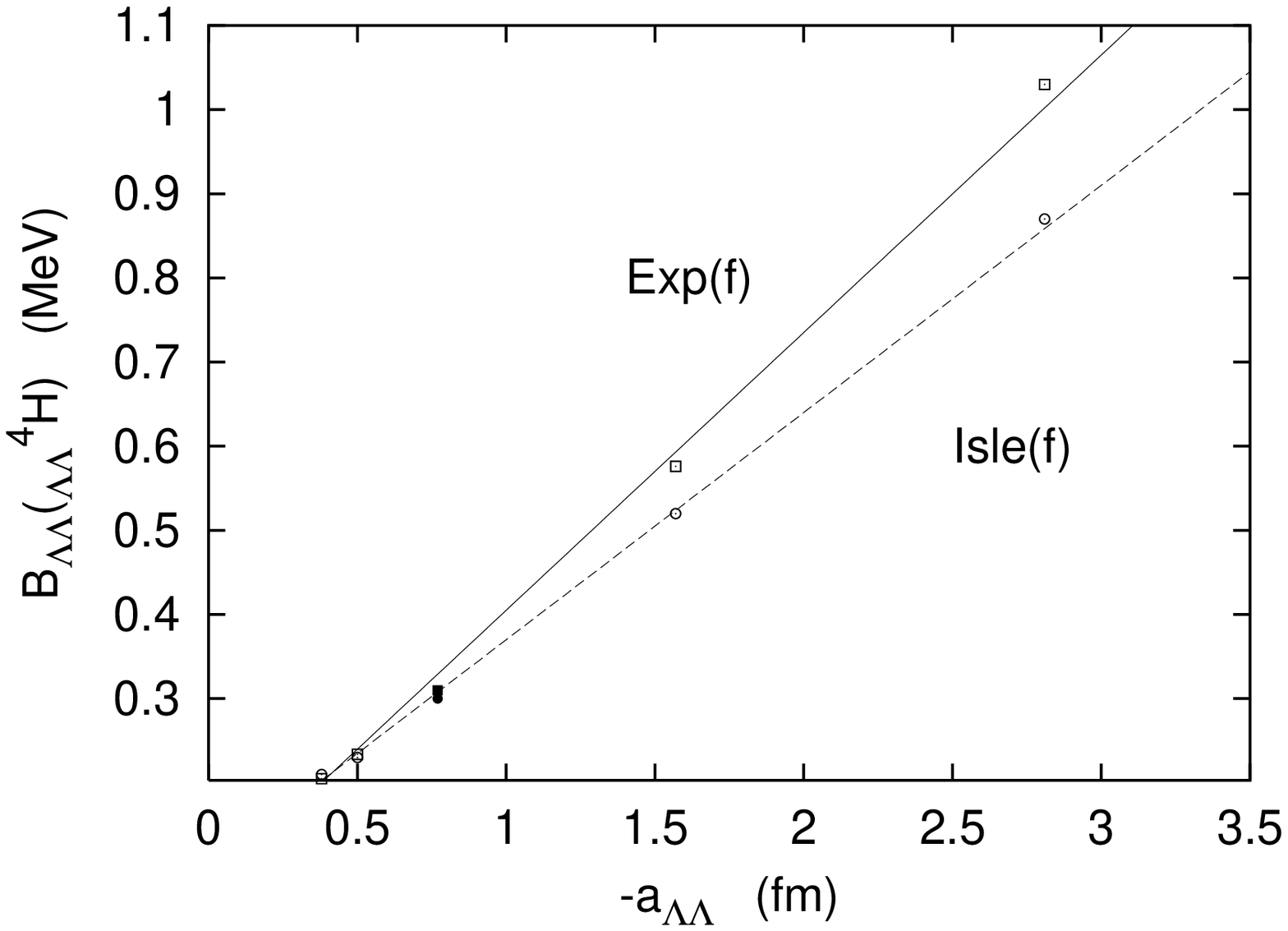,height=80mm,width=75mm} 
\caption{$s$-wave Faddeev calculations of 
$B_{\Lambda\Lambda}(_{\Lambda\Lambda}^{~~4}\rm H)$ 
in a $\Lambda\Lambda d$ model \cite{FGa02c}.} 
\label{fig:Isle} 
\end{minipage} 
\end{figure}

\subsection{$^{~~4}_{\Lambda\Lambda}${\rm H}} 

I start by discussing the first Faddeev-Yakubovsky (FY) four-body 
calculation of $^{~~4}_{\Lambda\Lambda}$H \cite{FGa02c}. 
For two identical hyperons and two essentially identical nucleons 
(upon introducing isospin) as appropriate to a $\Lambda\Lambda pn$ 
model calculation  of $^{~~4}_{\Lambda\Lambda}$H, the 18 FY 
components reduce to seven independent components satisfying coupled 
equations. Six rearrangement channels are involved in the $s$-wave 
calculation 
\cite{FGa02c} for $^{~~4}_{\Lambda\Lambda}$H(1$^+$): 
\begin{equation} 
(\Lambda NN)_{S=\frac12} + \Lambda \;, \;\; 
(\Lambda NN)_{S=\frac32} + \Lambda \;, \;\; 
(\Lambda \Lambda N)_{S=\frac12} + N \; 
\label{eq:3+1} 
\end{equation} 
for 3+1 breakup clusters, and 
\begin{equation} 
(\Lambda \Lambda)_{S=0} + (NN)_{S=1} \;, \;\;\; 
(\Lambda N)_S + (\Lambda N)_{S'} 
\label{eq:2+2} 
\end{equation} 
with $(S,S')$=$(0,1)$+$(1,0)$ and $(1,1)$ for the latter 2+2 
breakup clusters. 

\begin{figure} 
\begin{minipage}[t]{75mm} 
\epsfig{file=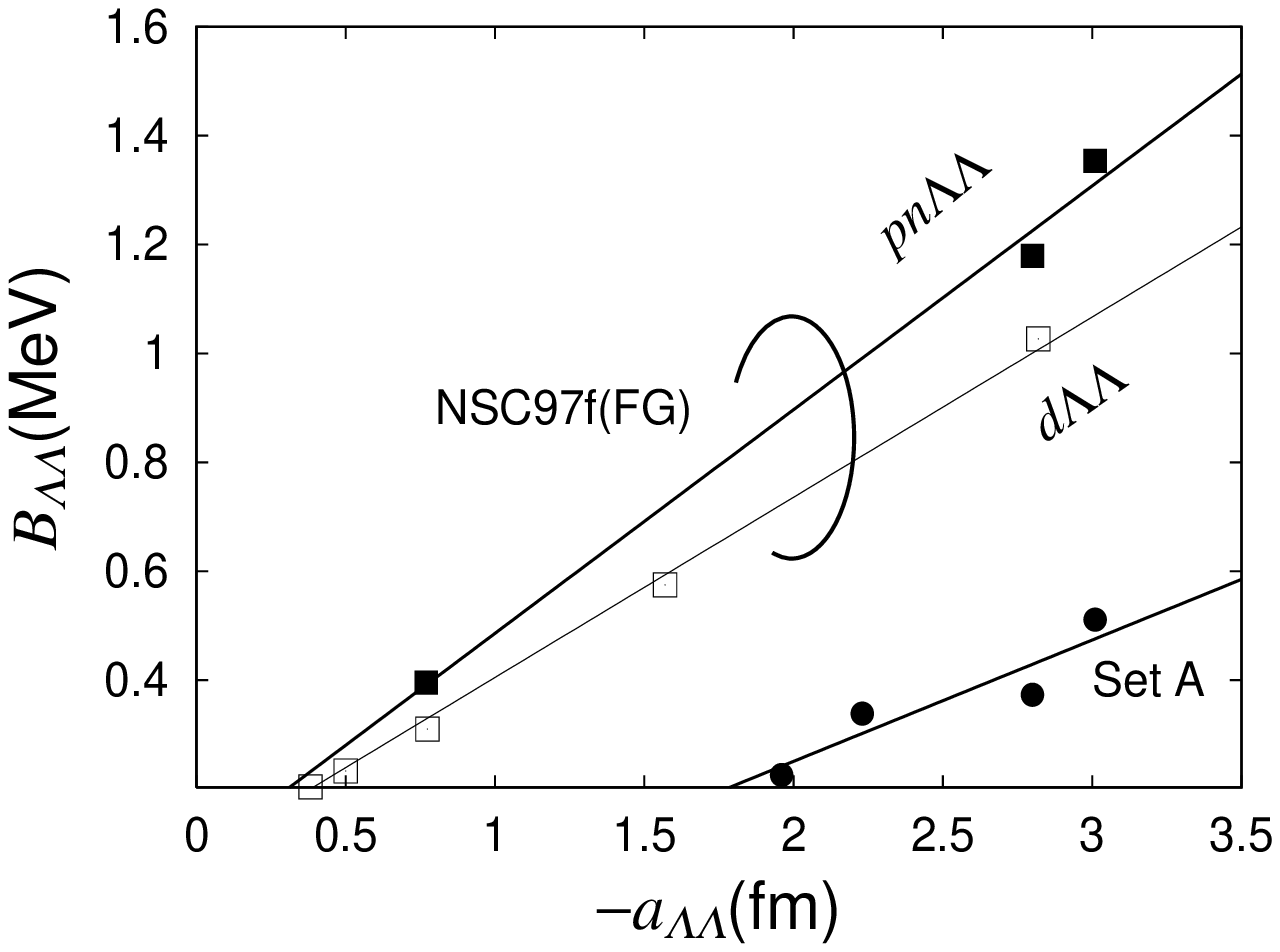,height=80mm,width=75mm} 
\caption{$B_{\Lambda\Lambda}(^{~~4}_{\Lambda\Lambda}$H) calculated 
by Nemura et al. \cite{NAM03} using the stochastic variational method.} 
\label{fig:akaishi} 
\end{minipage} 
\hspace{\fill} 
\begin{minipage}[t]{75mm} 
\epsfig{file=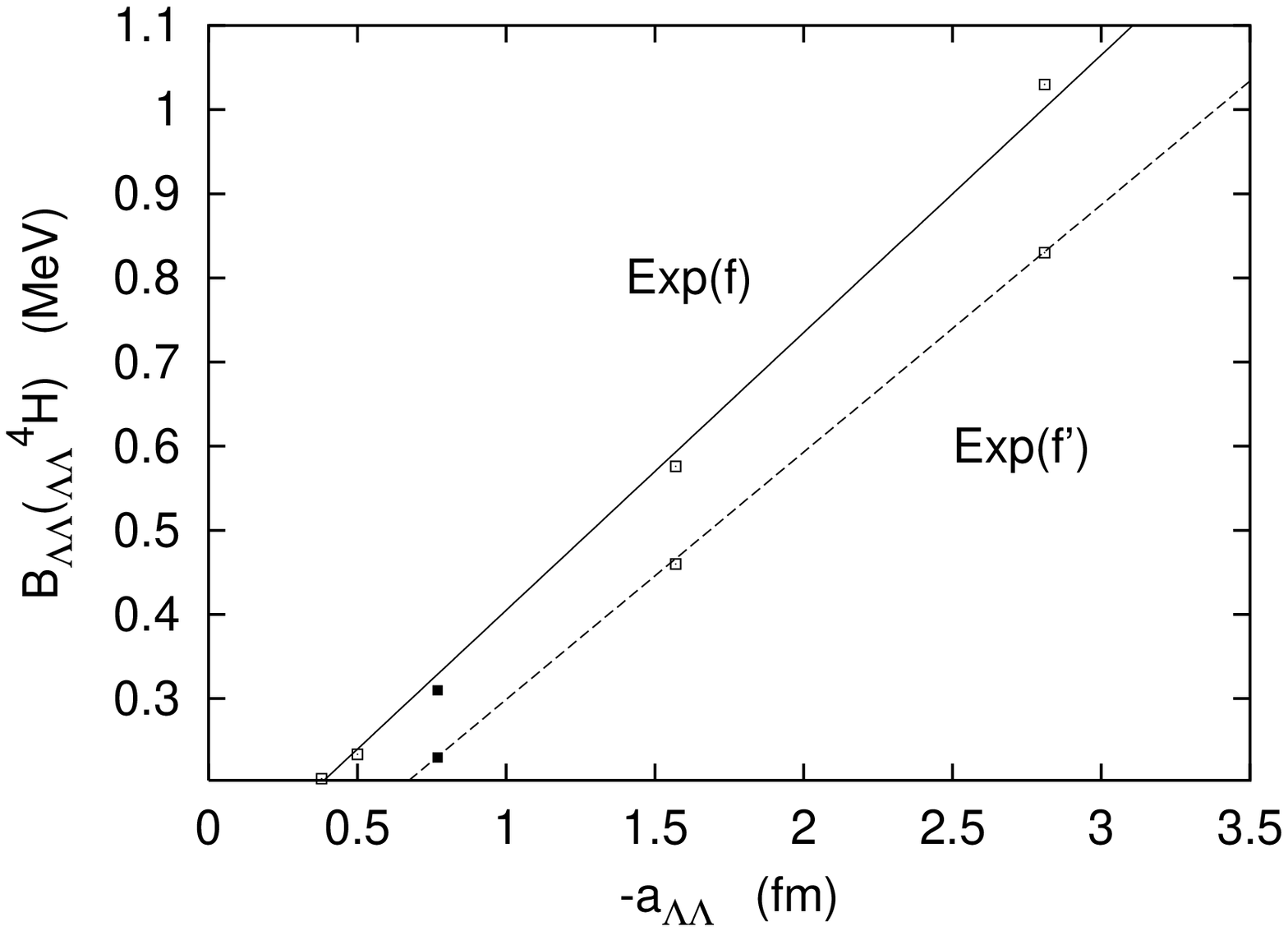,height=80mm,width=75mm} 
\caption{$B_{\Lambda\Lambda}(^{~~4}_{\Lambda\Lambda}$H) calculated 
in a $\Lambda\Lambda d$ model, using exponential $\Lambda d$ potentials 
constrained by NSC97f and NSC97f' \cite{FGa02c}.} 
\label{fig:Exp} 
\end{minipage} 
\end{figure} 

Using $V_{\Lambda\Lambda}$ which reproduces 
$B_{\Lambda\Lambda}(^{~~6}_{\Lambda\Lambda}$He), 
the four-body calculation converges well as function of the number 
$N$ of the FY basis functions allowed in, 
yet it yields no bound state for the $\Lambda\Lambda pn$ 
system, as demonstrated in Fig.\ref{fig:BE} by the location of the 
`$\Lambda\Lambda pn$' curve {\it above} the horizontal straight line 
marking the `$\Lambda + ~_{\Lambda}^{3}{\rm H}$ threshold'.\footnote{This 
threshold was obtained as the asymptote of the $\Lambda pn$ 
$s$-wave Faddeev calculation which uses model NSC97f \cite {RSY99} 
for the underlying $\Lambda N$ interaction, yielding 
$B_{\Lambda}(_{\Lambda}^3{\rm H}({\frac{1}{2}}^+))=0.19$ MeV. Using model 
NSC97e, with $B_{\Lambda}(_{\Lambda}^3{\rm H}({\frac{1}{2}}^+))=0.07$ MeV, 
does not alter the conclusions listed below.} 
In fact these FY calculations exhibit little sensitivity 
to $V_{\Lambda\Lambda}$ over a wide range. 
Even for considerably stronger $\Lambda\Lambda$ interactions one gets a bound 
$^{~~4}_{\Lambda\Lambda}$H only if the $\Lambda N$ interaction is made 
considerably stronger, by as much as 40\%. With four $\Lambda N$ pairwise 
interactions out of a total of six, the strength of 
the $\Lambda N$ interaction (about half of that for $NN$) plays a major 
role in the four-body $\Lambda\Lambda pn$ problem. 
However, fitting a $\Lambda d$ potential to the low-energy parameters 
of the $s$-wave Faddeev calculation for $\Lambda pn$ and solving the 
$s$-wave Faddeev equations for a $\Lambda\Lambda d$ model of 
$^{~~4}_{\Lambda\Lambda}$H, this latter four-body system is calculated to 
yield a $1^+$ bound state, as shown in the figure by the location of 
the asymptote of the `$\Lambda\Lambda d$' curve {\it below} the 
`$\Lambda + ~_{\Lambda}^{3}{\rm H}$ threshold'. 
The onset of particle stability for $^{~~4}_{\Lambda\Lambda}$H($1^+$) 
requires then a minimum strength for $V_{\Lambda\Lambda}$ 
which is exceeded by the choice 
of $B_{\Lambda\Lambda}(^{~~6}_{\Lambda\Lambda}$He) \cite{Tak01} 
as a normalizing datum 
(equivalent to $-a_{\Lambda\Lambda} \sim 0.8$ fm \cite{FGa02b}). 
This is demonstrated in Fig.\ref{fig:Isle} 
where Faddeev-calculated $B_{\Lambda\Lambda}(^{~~4}_{\Lambda\Lambda}$H) 
values are shown as function of the $\Lambda\Lambda$ scattering length 
$a_{\Lambda\Lambda}$ for two different functional forms of the fitted 
$\Lambda d$ potential. 
Disregarding spin it can be shown that, 
for essentially attractive $\Lambda\Lambda$ interactions and 
for a static nuclear core $d$, a two-body $\Lambda d$ bound state implies 
binding for the three-body $\Lambda\Lambda d$ system \cite{Bas02}. 
However, for a {\it non static} nuclear core $d$ 
(made out of dynamically interacting proton and neutron), 
a $\Lambda d$ bound state does not necessarily imply binding 
for the $\Lambda\Lambda d$ system. 
It is questionable whether by incorporating higher partial waves, or 
$\Lambda\Lambda - \Xi N$ coupling effects, this qualitative 
feature will change. 

\begin{figure} 
\begin{minipage}[t]{75mm} 
\epsfig{file=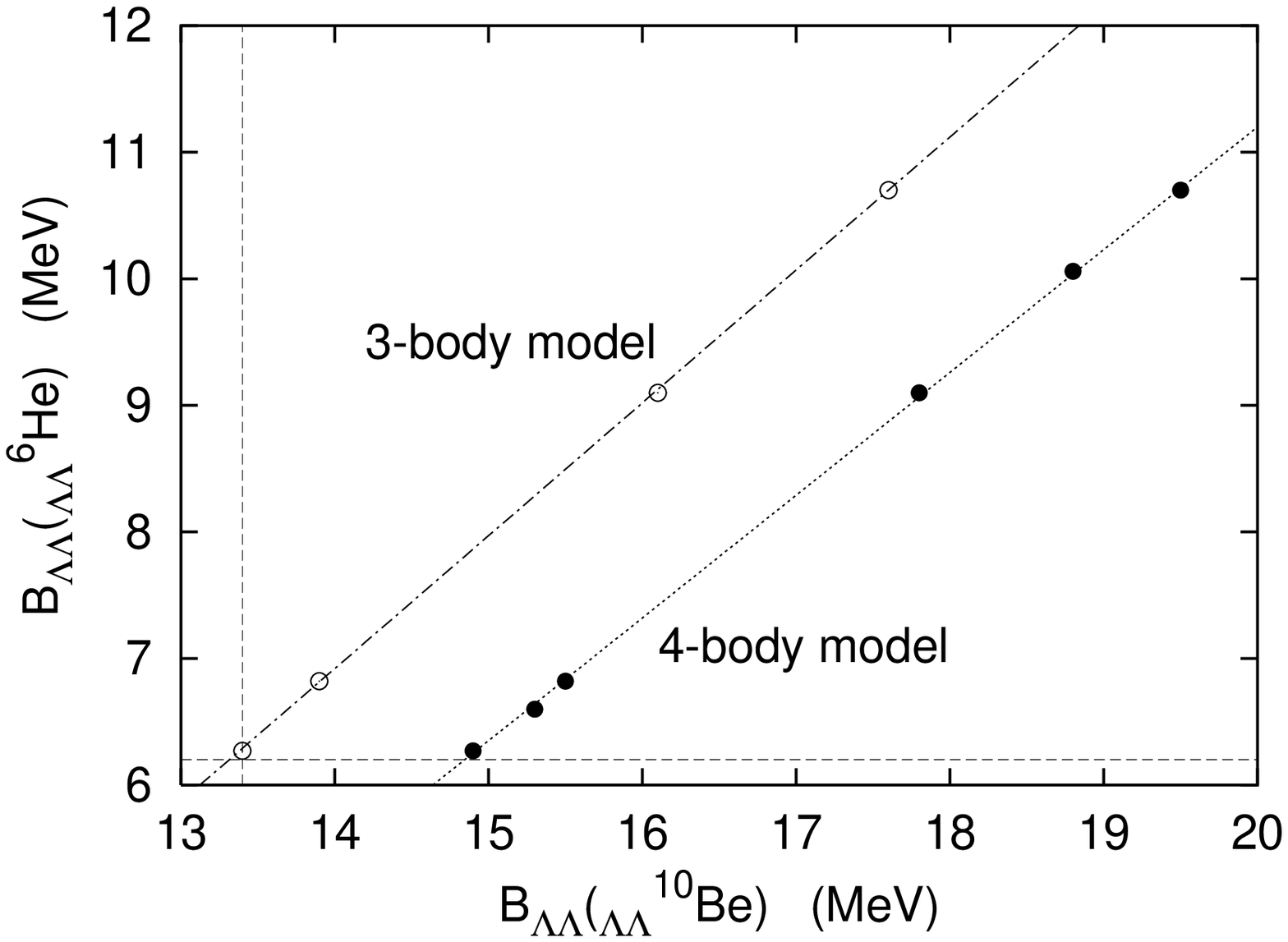,height=80mm,width=75mm} 
\caption{$s$-wave FY calculations \cite{FGa02b} for 
$^{~10}_{\Lambda\Lambda}$Be: $^8$Be$\Lambda \Lambda$ vs. 
$\alpha\alpha\Lambda\Lambda$.} 
\label{fig:e6e10} 
\end{minipage} 
\hspace{\fill} 
\begin{minipage}[t]{75mm} 
\epsfig{file=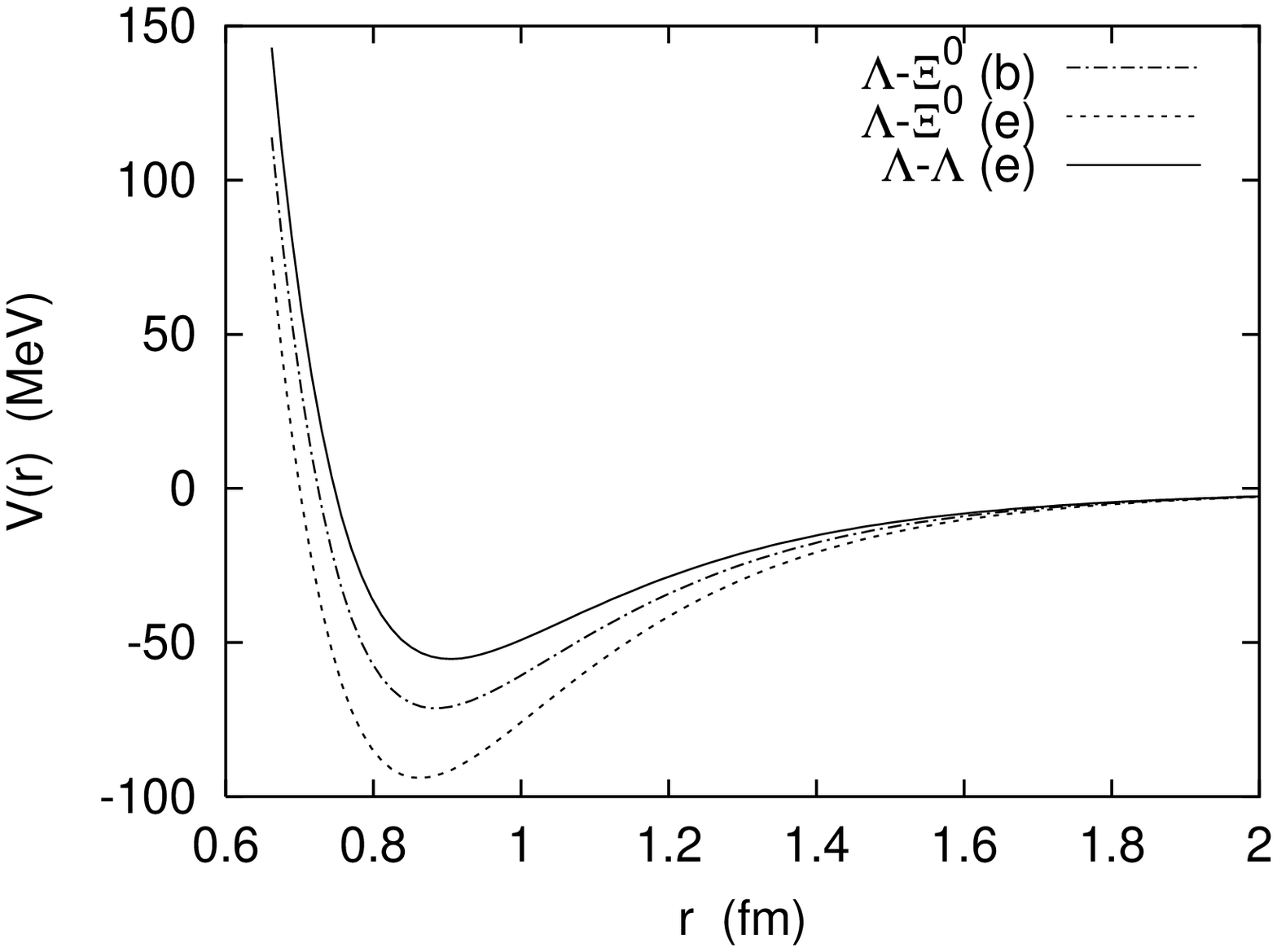,height=80mm,width=75mm} 
\caption{NSC97 phase-equivalent $YY$ potentials \cite{FGa02a}.} 
\label{fig:pot-LY} 
\end{minipage} 
\end{figure} 

The above conclusions have been very recently challenged by Nemura 
et al.\cite{NAM03}. Fig.\ref{fig:akaishi} demonstrates that 
within their stochastic variational calculation, which uses the 
NSC97f input of the Filikhin and Gal calculation \cite{FGa02c} 
for the various pairwise interactions, a `$pn\Lambda\Lambda$' model 
{\it always} yields more binding than a `$d\Lambda\Lambda$' model does. 
Particle stability for $^{~~4}_{\Lambda\Lambda}$H($1^+$) in this 
variational calculation requires a minimum strength for 
$V_{\Lambda\Lambda}$ which is exceeded by the choice 
of $B_{\Lambda\Lambda}(^{~~6}_{\Lambda\Lambda}$He) \cite{Tak01} 
as a normalizing datum. Yet, Nemura et al. argue that 
the $\Lambda N$ interaction in the $^3S$ channel, when adjusted to the 
binding energy calculated for the $A=4$ $\Lambda$ hypernuclei, should be 
taken weaker than that used by Filikhin and Gal and that, when this 
constraint is implemented (`set A' in the figure), particle stability for 
$^{~~4}_{\Lambda\Lambda}$H($1^+$) requires a minimum strength for 
$V_{\Lambda\Lambda}$ which is not satisfied by the choice 
of $B_{\Lambda\Lambda}(^{~~6}_{\Lambda\Lambda}$He) 
as a normalizing datum. A similar strong dependence on the $^3S$ 
$\Lambda N$ interaction within a $\Lambda\Lambda d$ model is shown 
in Fig.\ref{fig:Exp} for two versions $f$ and $f'$ of model NSC97 
which produce the same $B_{\Lambda}(_{\Lambda}^3{\rm H}({\frac{1}{2}}^+))$, 
while differing slightly by the location of the spin-flip 
${\frac{3}{2}}^+$ excited state \cite{FGa02c}.

\subsection{$^{~10}_{\Lambda\Lambda}${\rm Be}} 

For heavier $\Lambda\Lambda$ hypernuclei, 
the relationship between the three-body and four-body models is opposite 
to that found by Filikhin and Gal for $^{~~4}_{\Lambda\Lambda}$H: 
the $\Lambda\Lambda C_1 C_2$ calculation provides {\it higher} binding 
than a properly defined $\Lambda\Lambda C$ calculation yields 
(with $C=C_1+C_2$) due to the attraction induced by 
the $\Lambda C_1$-$\Lambda C_2$, $\Lambda\Lambda C_1$-$C_2$, 
$C_1$-$\Lambda\Lambda C_2$ four-body rearrangement channels that include 
bound states for which there is no room in the three-body $\Lambda\Lambda C$ 
model. The binding energy calculated within the four-body model increases 
then `normally' with the strength of $V_{\Lambda\Lambda}$ \cite{FGa02b}. 
This is demonstrated in Fig. \ref{fig:e6e10} for $^{~10}_{\Lambda\Lambda}$Be 
using several $\Lambda\Lambda$ interactions, including $V_{\Lambda\Lambda}=0$ 
which corresponds to the lowest point on each one of the straight lines. 
The origin of the dashed axes corresponds to $\Delta B_{\Lambda\Lambda} = 0$. 
Within the 4-body $\alpha\alpha\Lambda\Lambda$ model, the fairly large 
value $\Delta B_{\Lambda\Lambda}(^{~10}_{\Lambda\Lambda} \rm{Be}) \sim 1.5$ 
MeV in the limit $V_{\Lambda\Lambda} \rightarrow 0$ is due to the special 
$\alpha\alpha$ cluster structure of the $^8$Be core. 
The correlation noted in the figure 
between $^{~10}_{\Lambda\Lambda}$Be and $^{~~6}_{\Lambda\Lambda}$He 
calculations, and the consistency between various reports on their 
$B_{\Lambda\Lambda}$ values, are discussed by Filikhin and Gal 
\cite{FGa02b,FGa02a}. In particular, the two solid points next to the 
lowest one on the `4-body model' line in Fig. \ref{fig:e6e10}, 
corresponding to two versions of model NSC97 \cite{SRi99}, are close 
to reproducing (the `new') $B_{\Lambda\Lambda}(^{~~6}_{\Lambda\Lambda}$He) 
but are short of reproducing (the `old') 
$B_{\Lambda\Lambda}(^{~10}_{\Lambda\Lambda}$Be) by about $2.3 \pm 0.4$. 
This apparent discrepancy may be substantially reduced by accepting 
a $^{~10}_{\Lambda\Lambda}$Be weak decay scheme that involves the 3 MeV 
excited $^{9}_{\Lambda}$Be doublet rather than the $^{9}_{\Lambda}$Be 
ground state \cite{Dav03}. This conclusion may also be inferred from the 
recent 4-body calculations by Hiyama et al. for $A=7-10$ $\Lambda\Lambda$ 
hypernuclei \cite{HKM02}. 

\begin{figure} 
\epsfig{file=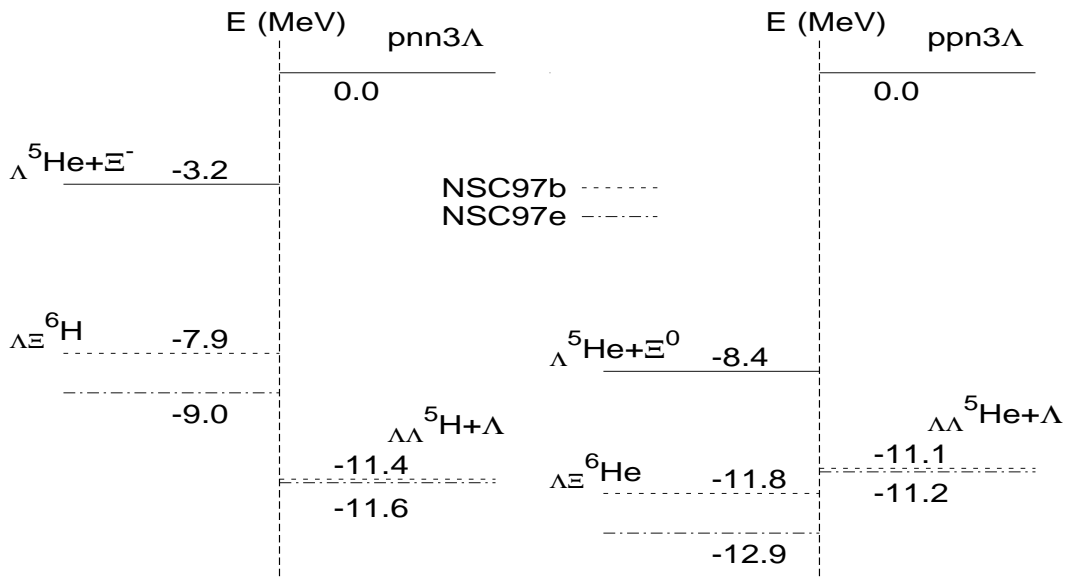,height=75mm,width=140mm} 
\caption{$s$-wave Faddeev calculations \cite{FGa02a} for 
$_{\Lambda\Xi}^{~~6}$H and $_{\Lambda\Xi}^{~~6}$He.} 
\label{fig:LXi} 
\end{figure}

\section{THE ONSET OF $\Xi$ STABILITY} 

Since model NSC97 \cite{RSY99} provides a qualitatively successful 
extrapolation from fits to $NN$ and $YN$ data to $S=-2$, and noting the 
strongly attractive $^{1}S_{0}$ $\Lambda\Xi$ potentials shown in 
Fig. \ref{fig:pot-LY} in comparison to the fairly weak $\Lambda\Lambda$ 
potentials when model NSC97 is extrapolated to $S=-3,-4$ \cite{SRi99}, 
it is natural to search for stability of $A=6, S=-3$ systems 
obtained from $^{~~6}_{\Lambda\Lambda}$He upon replacing one of the 
$\Lambda$'s by $\Xi$. Faddeev calculations \cite{FGa02a} 
for the $0^+$ $I=1/2$ ground-state of $^{~~6}_{\Lambda\Xi}$H and 
$^{~~6}_{\Lambda\Xi}$He, considered as $\alpha\Lambda\Xi^-$ and 
$\alpha\Lambda\Xi^0$ three-body systems respectively, 
indicate that $^{~~6}_{\Lambda\Xi}$He is particle-stable 
against $\Lambda$ emission to $^{~~5}_{\Lambda\Lambda}$He 
for potentials simulating model NSC97, 
particularly versions $e$ and $f$, 
whereas $_{\Lambda\Xi}^{~~6}$H is unstable since 
$M(\Xi^-) > M(\Xi^0)$ by 6.5 MeV.\footnote{Recall that the $I=1/2$ 
$^{~~5}_{\Lambda\Lambda}$H - $^{~~5}_{\Lambda\Lambda}$He 
hypernuclei, within a $\Lambda\Lambda C$ Faddeev calculation, 
are particle stable even in the limit $V_{\Lambda\Lambda} \rightarrow 0$.} 
This is demonstrated in Fig. \ref{fig:LXi}. Nevertheless, predicting 
particle stability for $^{~~6}_{\Lambda\Xi}$He is not independent 
of the assumptions made on the experimentally unexplored $\Xi \alpha$ 
interaction which was extrapolated from recent data on $^{12}$C \cite{Fuk98}; 
hence this prediction cannot be considered conclusive.

\section{STRANGE HADRONIC MATTER}

Bodmer \cite{Bod71}, and more specifically Witten \cite{Wit84}, suggested 
that strange quark matter, with roughly equal composition of $u$, $d$ and 
$s$ quarks, might provide the absolutely 
stable form of matter. Metastable strange quark matter has been studied by 
Chin and Kerman \cite{CKe79}. Jaffe and collaborators \cite{Jaf84,Jaf87} 
subsequently charted the various scenarios possible for the stability 
of strange quark matter, from absolute stability down to metastability 
due to weak decays. Finite strange quark systems, so called strangelets, 
have also been considered \cite{Jaf84,Jaf93}. 

Less advertised, perhaps, is the observation made by 
Schaffner et al. that metastable strange systems with similar properties, 
i.e. a strangeness fraction $f_{S} = -S/A \approx 1$ and a charge fraction 
$f_{Q} = Z/A \approx 0$, might also exist in the hadronic basis at moderate 
values of density, between twice and three times nuclear matter density 
\cite{SDG93,SDG94}. These strange systems are made out of $N$, $\Lambda$ 
and $\Xi$ baryons. The metastability of these strange 
hadronic systems was established by extending relativistic mean field 
(RMF) calculations from ordinary nuclei ($f_{S} = 0$) to 
multi-strange nuclei with $f_{S}\not= 0$. 
Although the detailed pattern of metastability, as well as 
the actual values of the binding energy, depend specifically on 
the partly unknown hyperon potentials assumed in dense matter, the 
predicted phenomenon of metastability turned out to be robust in 
these calculations \cite{BGS94}. 

Recently, model NSC97 and its extension \cite{SRi99} were used 
to calculate within the RMF framework the minimum-energy 
equilibrium composition of bulk strange hadronic matter (SHM) made out 
of the SU(3) octet baryons $N, \Lambda, \Sigma$ and $\Xi$, 
over the entire range of strangeness fraction $0 \leq f_{S} \leq 2$ 
\cite{SBG00}. The main result is that SHM is comfortably metastable 
in this model for any allowed value of $f_{S} > 0$. The 
$N \Lambda \Xi$ composition and the binding energy calculated 
for equilibrium configurations with $f_{S} \leq 1$ 
resemble those of model 2 in Refs. \cite{SDG93,SDG94}. 
The extension of model NSC97 \cite{SRi99} yields particularly attractive 
$\Xi \Xi$, $\Sigma \Sigma$ and $\Sigma \Xi$ interactions, but fairly 
weak $\Lambda \Lambda$ and $N \Xi$ interactions. 
Consequently, for $f_{S} \geq 1$, $\Sigma$'s replace 
$\Lambda$'s due to their exceptionally strong attraction to 
$\Sigma$ and $\Xi$ hyperons. As is shown below, a first-order 
phase transition occurs from $ N \Lambda \Xi $ dominated 
matter for $f_{S} \leq 1$ to $ N \Sigma \Xi$ 
dominated matter for $f_{S} \geq 1$, with 
binding energies per baryon reaching as much as 80 MeV. 

\begin{figure} 
\begin{minipage}[t]{75mm} 
\epsfig{file=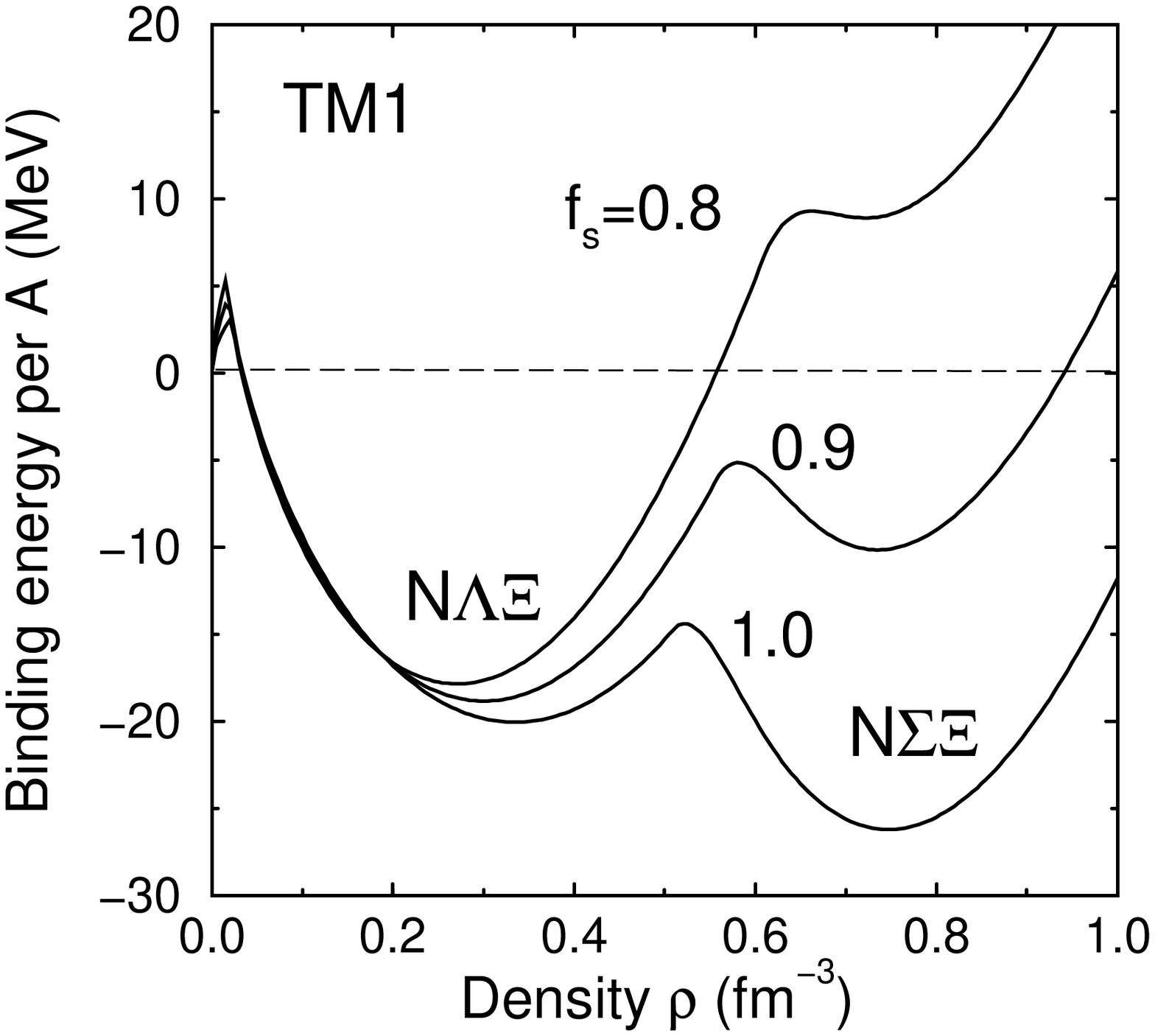,height=85mm,width=75mm} 
\caption{Transition from $N\Lambda\Xi$ to $N\Sigma\Xi$ matter 
upon increasing the strangeness fraction \cite{SBG00}.} 
\label{fig:SBG4} 
\end{minipage} 
\hspace{\fill} 
\begin{minipage}[t]{75mm} 
\epsfig{file=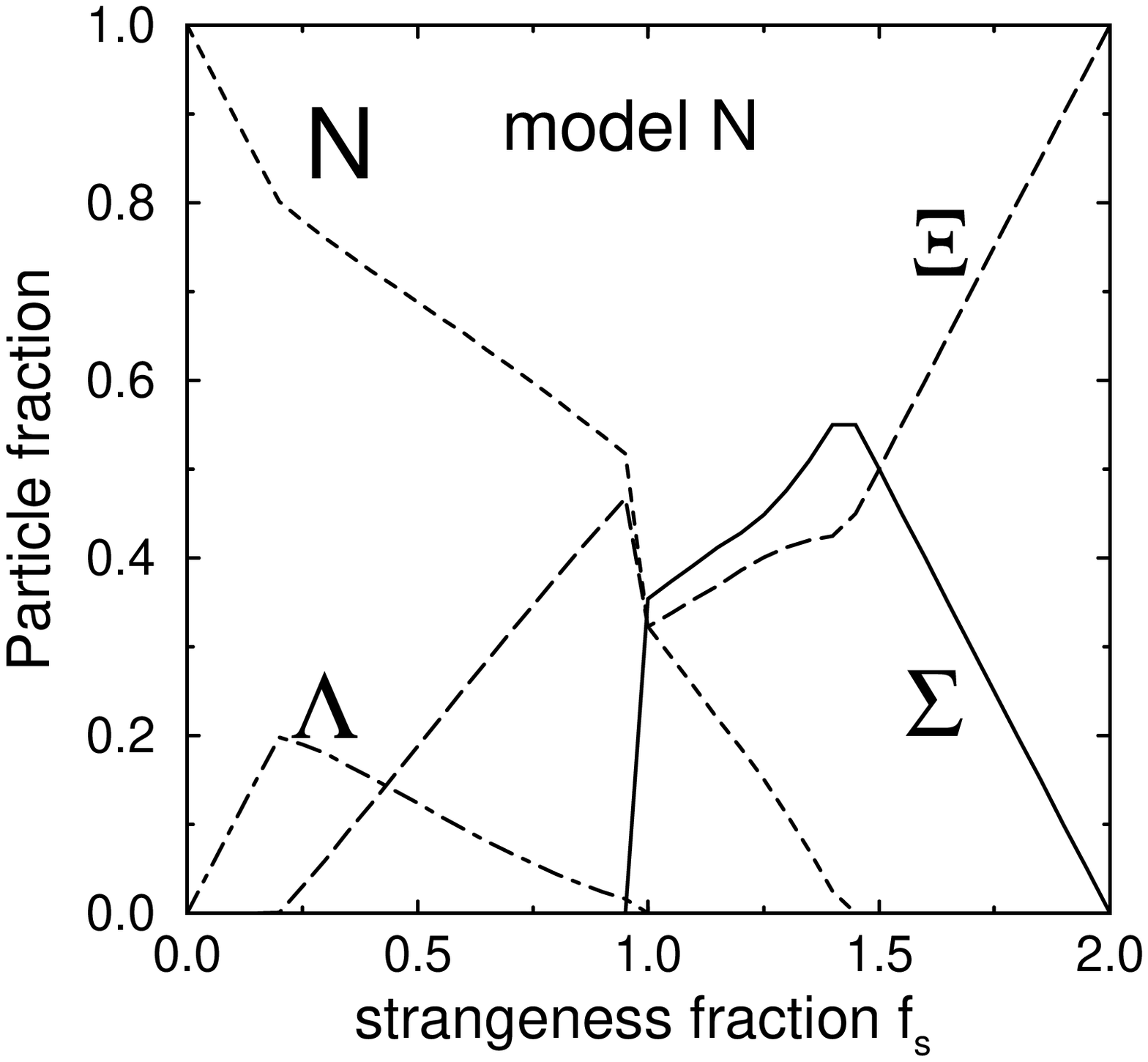,height=85mm,width=75mm} 
\caption{Strange hadronic matter composition as function of the 
strangeness fraction \cite{SBG00}.} 
\label{fig:SBG5} 
\end{minipage}  
\end{figure}

A phase transition is visualized in Fig. \ref{fig:SBG4} where the 
binding energy is drawn versus the baryon density 
for several representative fixed values of $f_S$. For $f_S=0.8$, 
there is a global minimum at a baryon density of $\rho_B=0.27$ fm$^{-3}$. 
A shallow local minimum is seen at larger baryon density at 
$\rho_B=0.72$ fm$^{-3}$. Increasing the strangeness fraction to $f_S=0.9$ 
lowers substantially the local minimum by about 20 MeV, whereas the global 
minimum barely changes. At $f_S=1.0$ this trend is amplified and the 
relationship between the two minima is reversed, as the minimum at higher 
baryon density becomes energetically favored. The system will then undergo 
a transition from the low density state to the high density state. 
Due to the barrier between the two minima, it is a first-order phase 
transition from one minimum to the other. 

Fig. \ref{fig:SBG5} demonstrates explicitly that the phase transition
involves transformation from $N\Lambda\Xi$ dominated matter to $N\Sigma\Xi$
dominated matter, by showing the calculated composition of SHM for this 
model (denoted N) as function of the strangeness fraction $f_S$.
The particle fractions for each baryon species change as function of
$f_S$. At $f_S=0$, one has pure nuclear matter, whereas at $f_S=2$ one has
pure $\Xi$ matter. In between, matter is composed of baryons as dictated by
chemical equilibrium. A change in the particle fraction may occur quite
drastically when new particles appear, or existing ones disappear in the
medium. A sudden change in the composition is seen in
Fig. \ref{fig:SBG5} for $f_S=0.2$ when $\Xi$'s emerge in the medium,
or at $f_S=1.45$ when nucleons disappear. The situation at $f_S=0.95$ is
a special one,  as $\Sigma$'s appear in the medium,
marking the first-order phase transition observed in the previous figure. The
baryon composition alters completely at that point, from $N\Xi$ baryons
plus a rapidly vanishing fraction of $\Lambda$'s into $\Sigma\Xi$ hyperons plus
a decreasing fraction of nucleons. At the very deep minimum of the binding 
energy curve (Fig. 3 of Ref. \cite{SBG00}) SHM is composed mainly of 
$\Sigma$'s and $\Xi$'s with a very small admixture of nucleons.

\section{CONCLUSION} 

I have presented Faddeev calculations for 
$^{~~5}_{\Lambda\Lambda}$H - $^{~~5}_{\Lambda\Lambda}$He and 
$^{~~6}_{\Lambda\Lambda}$He, and first ever four-body Faddeev-Yakubovsky 
calculations for $^{~~4}_{\Lambda\Lambda}$H and $^{~10}_{\Lambda\Lambda}$Be, 
using two-body potentials fitted to the low-energy scattering parameters 
or to the binding energies of the respective subsystems. In particular, 
for $^{~~4}_{\Lambda\Lambda}$H, $NN$ and $\Lambda N$ interaction 
potentials that fit the binding energy of $^3_\Lambda$H were used. 
No $^{~~4}_{\Lambda\Lambda}$H bound state was obtained for 
a wide range of $\Lambda\Lambda$ interactions, including that corresponding 
to $B_{\Lambda\Lambda}(^{~~6}_{\Lambda\Lambda}$He). 
This non binding is due to the relatively weak $\Lambda N$ 
interaction, in stark contrast to the results of a `reasonable' 
three-body $\Lambda\Lambda d$ Faddeev calculation. 
Further experimental work is needed to decide whether 
or not the events reported in the AGS experiment E906 \cite{Ahn01} 
correspond to $^{~~4}_{\Lambda\Lambda}$H, particularly in view of subsequent 
conflicting theoretical analyses \cite{KFO02,KKM03}. 
More theoretical work, particularly on the effects of including explicitly 
$\Lambda\Lambda - \Xi N - \Sigma\Sigma$ channel couplings, is called for. 
Preliminary estimates for such effects within the NSC97 model, or its 
simulation, have been recently made \cite{FGS03,AGi03,MSA03,VRP03,LYa03}. 
In particular, Lanskoy and Yamamoto \cite{LYa03} have focused attention to 
the substantial charge symmetry breaking effects introduced by the different 
$\Xi - ^4$He thresholds into the binding energy calculation of the 
$^{~~5}_{\Lambda\Lambda}$H - $^{~~5}_{\Lambda\Lambda}$He ground states. 
In addition to increasing the calculated $\Delta B_{\Lambda\Lambda}$ value 
of $^{~~5}_{\Lambda\Lambda}$He with respect to that of 
$^{~~5}_{\Lambda\Lambda}$H, on top of the difference already shown in 
Fig. \ref{fig:e6e5}, it is found that for model NSC97 
$\Delta B_{\Lambda\Lambda}(^{~~5}_{\Lambda\Lambda}$He) gets as large 
and perhaps even larger than that of $^{~~6}_{\Lambda\Lambda}$He 
($\sim 1$ MeV). A simultaneous determination of the binding energies of 
$^{~~5}_{\Lambda\Lambda}$He and $^{~~6}_{\Lambda\Lambda}$He would help 
to discriminate between several versions of OBE models which differ markedly 
from each other regarding the strength of the off-diagonal 
$\Lambda\Lambda - \Xi N$ coupling. 

Accepting the predictive power of model NSC97, Faddeev 
calculations suggest that $_{\Lambda\Xi}^{~~6}$He may be the lightest 
particle-stable $S=-3$ hypernucleus, and the lightest and least strange 
particle-stable hypernucleus in which a $\Xi$ hyperon is bound. 
Unfortunately, the direct production of $\Lambda\Xi$ hypernuclei 
is beyond present experimental capabilities, requiring the use of 
$\Omega^-$ initiated reactions. 

Finally, I have focused on the consequences of using model NSC97 for the 
binding and composition of strange hadronic matter. Strange hadronic matter 
is comfortably stable, up to weak decays, over a wide range of baryon-baryon 
interaction models, including model NSC97 here chosen because it 
successfully extrapolates from the $S=0,-1$ sectors in which 
it was constructed into the $S=-2$ sector, nearly reproducing 
$B_{\Lambda\Lambda}(^{~~6}_{\Lambda\Lambda}$He). The phase transition 
considered in this review has been recently discussed by the Frankfurt 
group \cite{SBG02} in the context of phase transition to hyperon matter 
in neutron stars. Unfortunately, it will take lots of imagination to devise 
experimentally a way to determine how attractive those 
$\Lambda\Xi$, $\Xi\Xi$, $\Xi\Sigma$, $\Sigma\Sigma$ interactions are, 
which are so crucial for the results exhibited in this review.

\end{document}